\documentclass[preprint,12pt]{elsarticle}
\usepackage{graphicx}
\usepackage{amssymb}

\journal{Solid State Communications}

\begin{document}

\begin{frontmatter}

\title{Magnetic properties of the ferrimagnetic cobaltite CaBaCo$_4$O$_7$}

\author[HMFL]{Zhe Qu\corref{cor1}}\ead{zhequ@hmfl.ac.cn}
\author[HMFL]{Langsheng Ling}
\author[HMFL]{Lei Zhang}
\author[USTC,HMFL]{Li Pi}
\author[HMFL,USTC]{Yuheng Zhang}
\address[HMFL]{High Magnetic Field Laboratory, Chinese Academy of Sciences, \\ Hefei, Anhui, 230031, China}
\address[USTC]{Hefei National Laboratory for Physical Sciences at the Microscale, \\ University of Science and Technology of China, Hefei, Anhui, 230026, China}
\cortext[cor1]{Corresponding author. Tel: +86-551-559-5640; Fax: +86-551-559-1149.}

\begin{abstract}
The magnetic properties of the ferrimagnetic cobaltite CaBaCo$_4$O$_7$ are systematically investigated. We find that the susceptibility exhibits a downward deviation below $\sim$ 360 K, suggesting the occurrence of short range magnetic correlations at temperature well above $T_C$. The effective moment is determined to be 4.5 $\mu_B$/f.u, which is consistent with that expected for the Co$^{2+}$/Co$^{3+}$ high spin species. Using a criterion given by Banerjee [Phys. Lett. \textbf{12}, 16 (1964)], we demonstrate that the paramagnetic to ferrimagnetic transition in CaBaCo$_4$O$_7$ has a first order character.
\end{abstract}


\end{frontmatter}

\section{Introduction}

Transition metal oxides with geometry frustration have attracted considerable interest over decades. \cite{GFM1,GFM2,GFM3,GFM4} They commonly exhibit the persistence of strong spin fluctuations at low temperatures. As a result, the long-range magnetic order is at least partially suppressed and various short range correlated phases such as spin liquid, spin glass or spin ice develop. In some cases, frustration can be partially or entirely released, either by structural distortions that lift the ground-state degeneracy, or by the "order-by-disorder" mechanism, \cite{liftGFM} resulting in the establishment of a long-range magnetic order.

\begin{figure}[t]
\includegraphics[angle=-90]{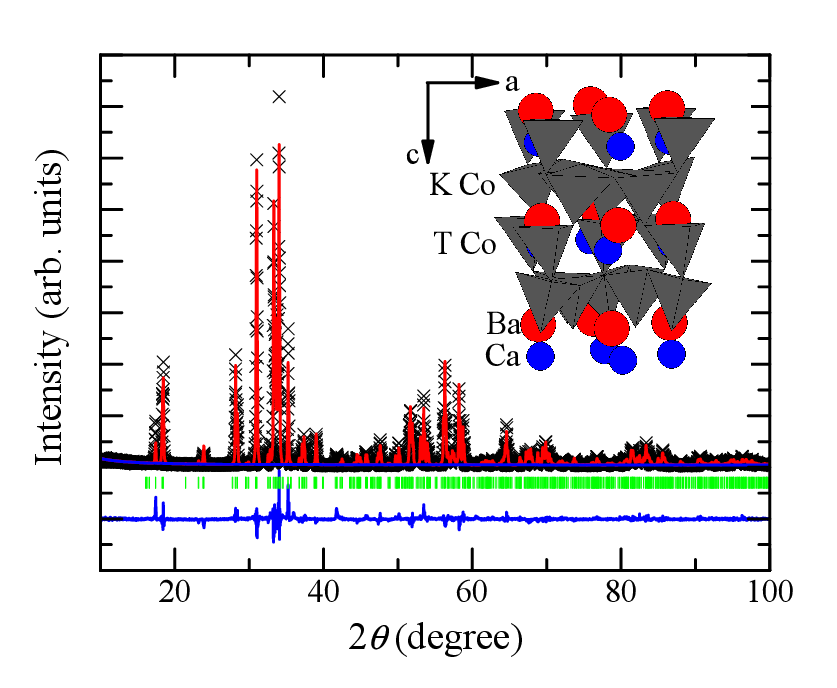}
\caption{(Color online) Powder XRD patterns of CaBaCo$_4$O$_7$. The solid curve is the best fit from the Rietveld refinement using GSAS, with $R_p =$ 11.68\% and $R_{wp} =$ 10.08\%. The vertical marks indicate the position of Bragg peaks and the bottom curves show the difference between the observed and calculated intensities. Inset shows the structure of CaBaCo$_4$O$_7$ viewed along $b$ axis. K Co and T Co represent kagome layer and triangular layer of CoO$_4$ tetrahedra, respectively.}\label{fig:XRD}
\end{figure}

Two well-known structural topology causing the presence of geometry frustration are two-dimensional triangular lattice and two-dimensional kagome lattice. Compositions whose structural motif embraces triangular or kagome layers are of great interest as model systems and have been the focus of numerous studies. In this respect, the recently discovered "114" cobaltite CaBaCo$_4$O$_7$ \cite{CaBaCo4O7SSC,CaBaCo4O7neutron} is particularly interesting because its crystal structure is built up of an alternate stacking of triangular or kagome layers formed by the CoO$_4$ tetrahedra (shown in the inset to Fig. \ref{fig:XRD}). There is very large distortion in the crystal, characterized by a strong buckling of the kagome layers. \cite{CaBaCo4O7SSC,CaBaCo4O7neutron} In addition, it exhibits charge ordering, with Co$^{2+}$ sitting on two sites and "mixed valent" cobalt Co$^{3+}$/Co$^{2+}L$ sitting on two other sites. \cite{CaBaCo4O7neutron} Due to the large structural distortion and the charge ordering, the geometry frustration is lifted, resulting in a ferrimagnetic ground state at low temperatures. \cite{CaBaCo4O7SSC,CaBaCo4O7neutron}

Although significant progress has been made in understanding the magnetic properties in CaBaCo$_4$O$_7$, a few questions remain to be answered. For example, does the system shows short-range magnetic correlations above $T_C$ like their "114" cousins such as YBaCo$_4$O$_7$? \cite{YBaCo4O7neutron1,YBaCo4O7neutron2} Why the obtained effective moment differs significantly from the expected value in CaBaCo$_4$O$_7$? \cite{CaBaCo4O7SSC} What's the nature of the paramagnetic to ferrimagnetic transition?

To address these questions, we systematically measured the magnetic properties of CaBaCo$_4$O$_7$. It is found that the susceptibility exhibits an downward deviation below $\sim$ 360 K, suggesting the occurrence of short range magnetic correlations at temperature well above $T_C$. By extending the magnetization measurement up to 800 K, the effective moment is determined to be 4.5 $\mu_B$/f.u through a Curie-Weiss analysis, which is consistent with that expected for the Co$^{2+}$/Co$^{3+}$ high spin species. Using a criterion given by Banerjee, \cite{BanerjeeCriteria} we demonstrate that the paramagnetic to ferrimagnetic phase transition in CaBaCo$_4$O$_7$ is a first order one.

\section{Experiment}

Polycrystalline sample of CaBaCo$_4$O$_7$ was prepared by using the conventional solid-state reaction method described in Ref. \cite{CaBaCo4O7SSC}. Stoichiometric proportions of high purity CaCO$_3$, BaCO$_3$ and Co$_3$O$_4$ were mixed and heated at 900 $^{o}C$ in air to decarbonation. They are then pelletized, and then sintered at 1100 $^{o}C$ in air for 12 hours and quenched to room temperature. The structure and the phase purity of the samples were checked by powder X-ray diffraction (XRD) at room temperature. Magnetization measurements were performed with a commercial superconducting quantum interference device (SQUID) magnetometer (Quantum Design MPMS 7T-XL) and a Physical Property Measurement System (Quantum Design PPMS-16T) equipped with a vibrating sample magnetometer (VSM).

\section{Results and Discussion}

Figure \ref{fig:XRD} displays the powder XRD pattern of CaBaCo$_4$O$_7$ at room temperature. Rietveld refinement \cite{GSAS,EXPGUI} of the XRD pattern confirms that the sample is single phase with an orthorhombic structure ($Pbn2_1$ space group). The lattice parameters are determined to be $a =$ 6.2871 ${\AA}$, $b =$ 11.0106 ${\AA}$ and $c =$ 10.1945 ${\AA}$, which agree well with previous reports within the experimental error. \cite{CaBaCo4O7SSC,CaBaCo4O7neutron}

\begin{figure}[t]
\includegraphics[angle=0]{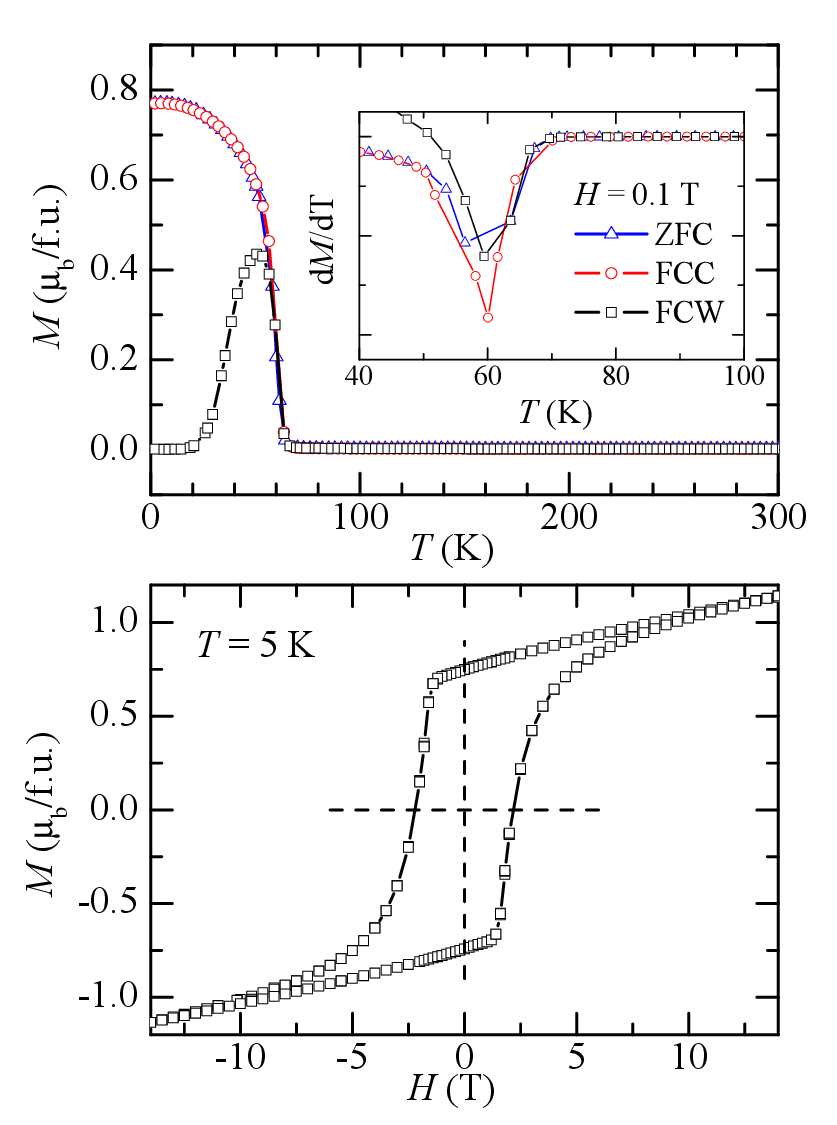}
\caption{(Color online) Upper panel: The magnetization as function of the temperature under an applied field of 0.1 T. inset shows the $dM/dT$ as function of the temperature. Lower panel: the magnetization as function of the field measured at 5 K.}\label{fig:M}
\end{figure}

The temperature dependence of the magnetization $M$($T$) between 2 K and 400 K under 0.1 T is shown in the upper panel of Fig. \ref{fig:M}. They are measured during field cooling sequence (FCC), during warming after field cooling sequence (FCW) and during warming after zero field cooling sequence (ZFC), respectively. All curve shows a rapid increase below $\sim$ 70 K, suggesting the occurrence of the transition into a magnetically ordered state. At 5 K, the saturated magnetic moment is still relatively small, only $\sim$ 1.1 $\mu_B/f.u.$ under 14 T (see the lower panel of Fig. \ref{fig:M}), agreeing with a ferrimagnetic ground state. The Curie temperature, defined as the temperature corresponding to the maximum in the $dM/dT$ curve, is determined to be 60 K (see the inset to Fig. \ref{fig:M}). Below $T_C$, we observe significant irreversibility between the magnetization curve measured after ZFC and FCC histories. This is attributed to the large coercive field compared to the applied field. \cite{CaBaCo4O7SSC,coercivity} As shown in the lower panel of Fig. \ref{fig:M}, the coercive field of CaBaCo$_4$O$_7$ is about 2 T at 5 K, which is much larger than the applied field of 0.1 T. Therefore, the magnetic domains will be locked in random direction during ZFC sequence while be aligned to the same direction during FCC or FCW sequences, resulting in the large irreversibility below $T_C$. All these results are consistent with previous report, \cite{CaBaCo4O7SSC} confirming that our sample is of high quality.

\begin{figure}[t]
\includegraphics[angle=-90]{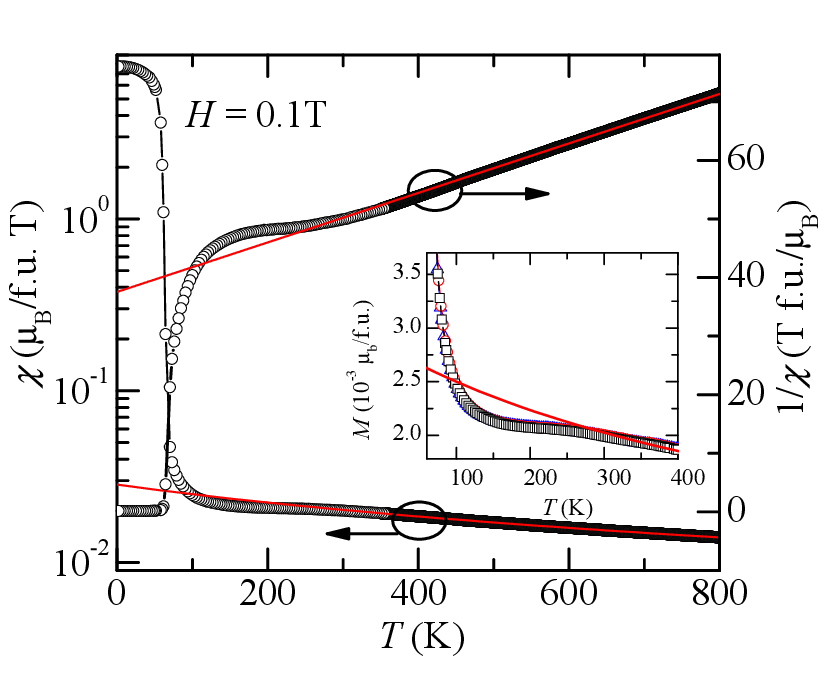}
\caption{(Color online) The susceptibility and the reciprocal of the susceptibility as function of temperature between 2 and 800 K under 0.1 T measured in FCC sequence. The solid lines represent the Curie-Weiss fitting. Inset shows the enlarged view of the $M(T)$ curve to highlight the deviation from the Curie-Weiss fitting.}\label{fig:CW}
\end{figure}

A close look on the temperature dependence of the magnetization reveals more information. The inset to Fig. \ref{fig:CW} displays the enlarged view of the $M(T)$ curves between $T_C$ and 400 K. One can see that while the magnetization decreases with increasing temperature above $T_C$ the slope of the $M(T)$ curve does not decrease monotonously as that expected for a pure paramagnetic state where the Curie-Weiss law predicts $\chi \propto C/(T-T_{CW})$ (Here $C$ is the Curie constant and $T_{CW}$ is the Curie-Weiss temperature). In order to understand this behavior, we extend the measurement of the $M$($T$) during FCC sequence up to 800 K and perform the Curie-Weiss analysis. As shown in Fig. \ref{fig:CW}, 1/$\chi$ shows an upward deviation from the linearity below $\sim$ 360 K. Since the deviation temperature is much higher than the Curie temperature, this behavior could not be attributed to the critical enhancement of the spin fluctuations due to the approach to the paramagnetic to ferrimagnetic transition but suggests the occurrence of short range magnetic correlations. The Curie-Weiss temperature $T_{CW}$ is determined to be $\sim$ -890 K. This gives $f = T_{CW}/T_C \sim 14.8$ which means CaBaCo$_4$O$_7$ is strongly frustrated. The effective moment is determined to be $\sim$ 4.5 $\mu_B/f.u.$, which agrees well with the value of a 1:1 combination of Co$^{2+}$/Co$^{3+}$ high spin species expected based on the chemical formula. The Curie-Weiss temperature and the effective moment obtained here are different from previous report. \cite{CaBaCo4O7SSC} This should be understood because short-range magnetic correlations might appears in their fitting temperature region.

\begin{figure}[t]
\includegraphics[angle=-90]{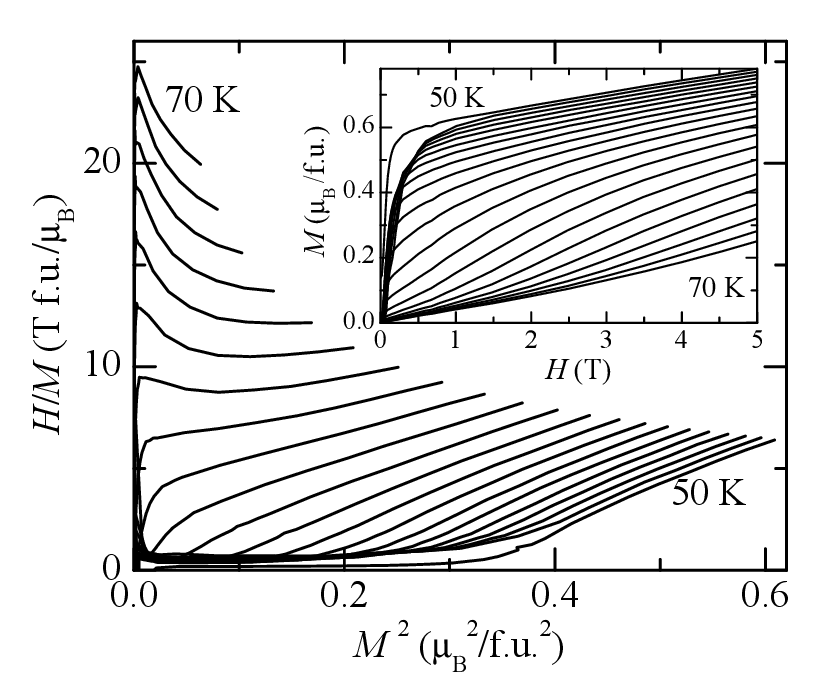}
\caption{Inset shows the initial isothermal magnetization curves at temperatures in the vicinity of the Curie temperature $T_C =$ 60 K at an interval of 1 K. The main panel shows these curves replotted as $H/M vs. M^2$.}\label{fig:MH}
\end{figure}

In order to obtain further information on the paramagnetic to ferrimagnetic transition in CaBaCo$_4$O$_7$, we use the criteria proposed by Banerjee to determine the order of this transition. By considering the essential similarity between the Landau-Lifshitz \cite{Landau-Lifshitz} and Bean-Rodbell \cite{Bean-Rodbell} criteria, Banerjee shows that the slope of the $H/M$ versus $M^2$ curves near the critical temperature can distinguish the first-order magnetic transition from the second order ones: a negative slope means the former and a positive slope means the latter. \cite{BanerjeeCriteria} We then measured the initial isothermal magnetization curves at temperatures in the vicinity of the Curie temperature. Before each run, the sample is warmed up to 200 K and then cooled to the measuring temperature under zero field to ensure a perfect demagnetization of the sample. The data are summarized in the inset to Fig. \ref{fig:MH}. It is noted that the $M(H)$ curve exhibits a peculiar behavior that its slope shows a decrease before an increase at intermediate fields. This behavior was also observed in MnAs, where a first order transition occurs at its Curie temperature and is used to test Banerjee's criteria. \cite{BanerjeeCriteria,Bean-Rodbell} We replotted the $M(H)$ curves as $H/M vs. M^2$ in Fig. \ref{fig:MH}. Negative slope is clearly observed between 64 and 70 K, which confirms that the paramagnetic to ferrimagnetic transition occurred in CaBaCo$_4$O$_7$ has a first order character according to the criterion.

\section{Conclusion}

In conclusion, we systematically investigate the magnetic properties of CaBaCo$_4$O$_7$. The Curie-Weiss temperature is determined to be $\sim$ -890 K and the effective moment be $\sim$ 4.5 $\mu_B/f.u.$. The susceptibility shows downward deviation from the Curie-Weiss law below $\sim$ 360 K, hinting that short range magnetic correlations might occur at temperature much higher than $T_C$ = 60 K. The paramagnetic to ferrimagnetic transition in CaBaCo$_4$O$_7$ is found to have the first order character.

\section{Acknowledgments}
This work is financially supported by the National Key Basic Research of China under Grant No. 2007CB925001 and 2010CB923403, and by National Natural Science Foundation of China under contract No. 11004198.

\newpage

\end{document}